\documentclass[3p,times,procedia]{elsarticle}
\usepackage{nupha_ecrc}

\volume{00}

\firstpage{1}

\journalname{Nuclear Physics A}

\runauth{B. Krouppa, A. Rothkopf, M. Strickland}

\jid{nupha}

\jnltitlelogo{Nuclear Physics A}

\usepackage{amssymb}

\usepackage[figuresright]{rotating}

\usepackage{amsmath}
\usepackage{natbib} 

\begin{document}

\begin{frontmatter}



\dochead{XXVIIth International Conference on Ultrarelativistic Nucleus-Nucleus Collisions\\ (Quark Matter 2018)}

\title{Bottomonium suppression at RHIC and LHC}


\author[Kent]{Brandon Krouppa}
\author[RK]{Alexander Rothkopf}
\author[Kent]{Michael Strickland}

\address[Kent]{Department of Physics, Kent State University, Kent, Ohio United States}
\address[RK]{Department of Mathematics and Physics, University of Stavanger, Stavanger, Norway}

\begin{abstract}
The strong suppression of heavy quarkonia is a good indicator that one has generated a quark-gluon plasma (QGP) in an ultrarelativistic heavy ion collision. Recent advancements in first principles calculations of the heavy quark potential provide additional insight into this suppression and can be used to further pin down the in-medium properties of the QGP. Realistic 3+1d dissipative hydrodynamical models can be used to simulate the QGP and, with bottomonium as a probe, one can make inferences for the initial temperature of the QGP and other phenomenological parameters such as the shear viscosity to entropy density ratio. However, progressing to LHC $\sqrt{s_{NN}}=5.02$ TeV Pb-Pb collisions, one expects regeneration to have an increasingly important impact on the suppression observables. In this proceedings, we present a brief overview to set up the model and then provide model results for bottomonium suppression and regeneration, for ultrarelativistic heavy-ion collision experiments at RHIC and LHC. 
\end{abstract}


\end{frontmatter}


\section{Introduction}
\label{sec:intro}

The ultrarelativistic heavy-ion collision (URHIC) experiments at the Relativistic Heavy Ion Collider (RHIC) at Brookhaven National Laboratory and the Large Hadron Collider (LHC) at CERN generate a hot medium of deconfined matter known as the quark-gluon plasma (QGP). Recent comparisons between theory and experiment suggest QGP formation with initial temperatures around $T_{0} = 600-700$ MeV at $\tau_{0}=0.25$ fm/c for $\sqrt{s_{NN}}=5.02$ TeV/nucleon Pb-Pb collisions at the LHC. Studies also suggest the QGP formed in URHICs behaves like a nearly-perfect fluid with a shear viscosity to entropy density ratio $\eta/s$ of $4\pi\eta/s \sim 1-3$. Bottomonium is of particular interest to study the QGP formed in URHICs due to several unique properties. With their relatively large vacuum binding energies ($\lesssim 1$ GeV), they can survive well into the phase of deconfined matter. Also, production yield ratios between the ground $\Upsilon(1S)$ and excited states reveal insight into properties of the QGP like the initial temperature and $\eta/s$.

The formation of bottomonium states before $\tau \sim 1$ fm/c mean bottomonia are sensitive to non-equilibrium early-time dynamics. Characterized by a breaking of symmetry of the $\mathcal{P}_{L}-\mathcal{P}_{T}$ plane, one can model to good approximation \cite{Song:2009gc}, a spheroidal form for the one-particle distribution function in the QGP which is due to anisotropies in momentum space. The initial geometry of a URHIC provides a system which is rapidly expanding along the longitudinal direction. After a short amount of time, the finite speed of sound in the QGP catches up and the shockwave pushes expansion in the transverse plane. Moreover, bottomonia are sensitive to this momentum-space anisotropy in the heavy quark potential calculation \cite{Dumitru:2007hy,Burnier:2009yu,Dumitru:2009fy,Strickland:2011mw,Strickland:2011aa}.

With increasing collision energies, e.g. at the LHC, one to expects a QGP which is hotter. With a broader inelastic cross section, the formation of $b\bar{b}$ pairs should become more common in URHICs. The increase in $b\bar{b}$ pairs provides a foundation for bottom quark and anti-bottom quark pairs to recombine following bottomonium state breakup, or combination after the pair forms from open bottom states. In this proceedings, we introduce a more comprehensive model for bottomonium suppression and regeneration at the RHIC and LHC experiments. 

\section{Model implementation and results}
\label{sec:results}
The model presented herein uses a 3+1d anisotropic hydrodynamics (aHydro) code that simulates the QGP, two different potentials \cite{Strickland:2011aa,Krouppa:2017jlg} which are used to describe the six bottomonium states' binding energy as a function of energy density and momentum-space anisotropy. A non-relativistic QCD (pNRQCD)-based potential (Strickland-Bazow) is used along with a new potential derived in large part from first principle lattice QCD studies. A regeneration model used \cite{BraunMunzinger:2000px,Du:2017qkv}, which is based on a first-order rate equation to compute the local particle density function,
\begin{equation}
\dfrac{dn(\tau, {\bf x})}{d\tau} = -\Gamma(T(\tau, {\bf x}))\Big[n(\tau, {\bf x})-n_{\text{eq}}(T(\tau, {\bf x}))\Big],
\end{equation}
where $n(\tau,{\bf x})$ is the bottomonium number density as a function of the spatiotemporal evolution of the QGP, $n_{\text{eq}}(\tau,{\bf x})$ is the equilibrium distribution of bottomonium, and $\Gamma(T(\tau,{\bf x}))$ is the temperature-dependent breakup rate which is primarily associated with the imaginary part of the bottomonium binding energy. Input for the regeneration model includes bottomonium state cross sections from experiment \cite{Du:2017qkv,Khachatryan:2016xxp,CMS:2017ucd,Patrignani:2016xqp,Abelev:2014qha,Aaij:2014caa}, and leads to modification of $n_{\text{eq}}$ due to the bottom quark fugacity $\gamma_{b}$. Initial temperatures for RHIC 200 GeV/nucleon collisions are $T_{0} = \{0.442,0.440,0.439\}$ and and for LHC 5.02 TeV collisions we have $T_{0} = \{0.641,0.632,0.629\}$ for $4\pi\eta/s=\{1,2,3\}$, respectively.

\begin{figure}[t!]
\centerline{
\includegraphics[width=0.502\linewidth, trim=0 0.13in 0 0]{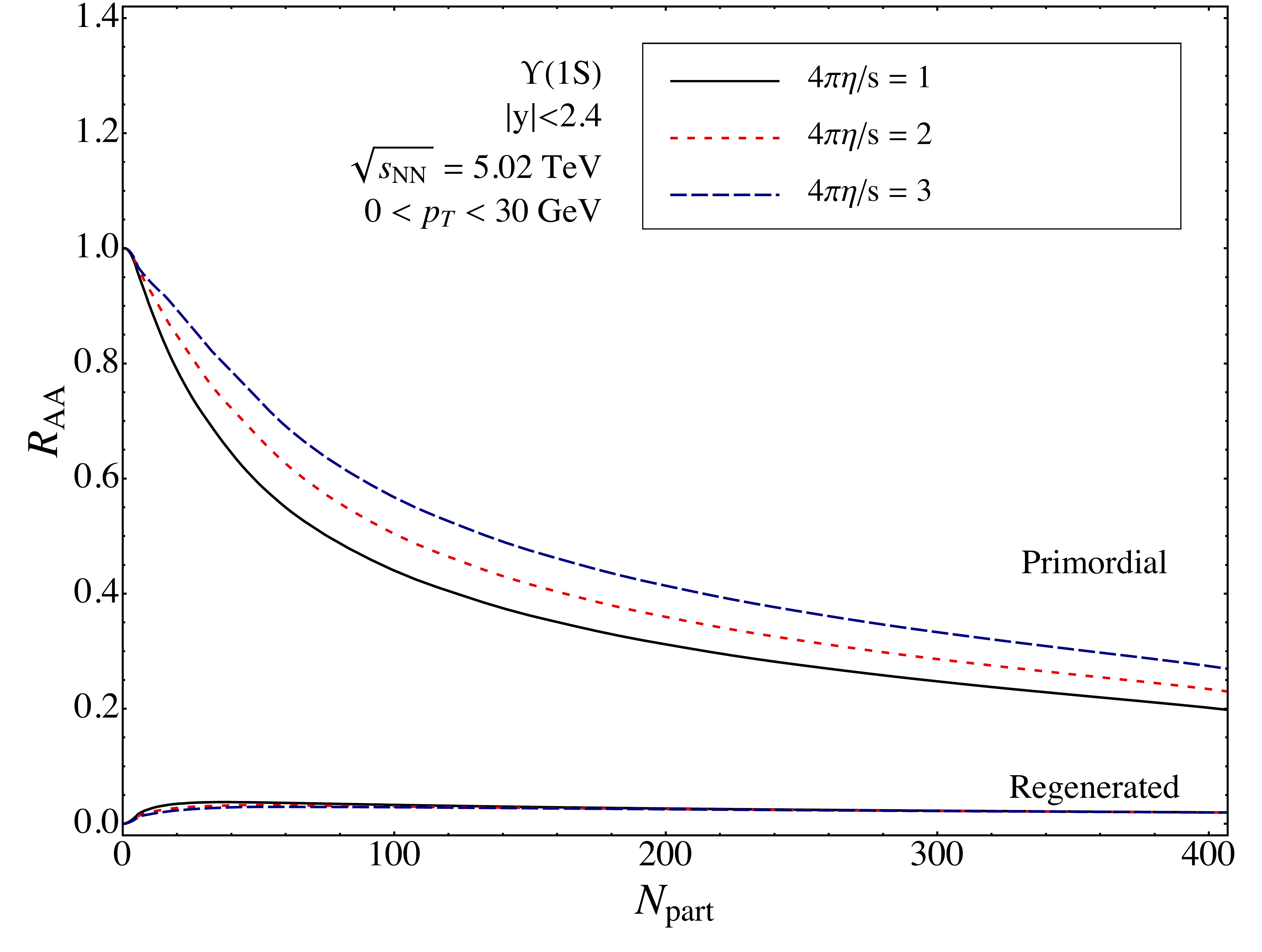}\hspace{2mm}
\includegraphics[width=0.485\linewidth]{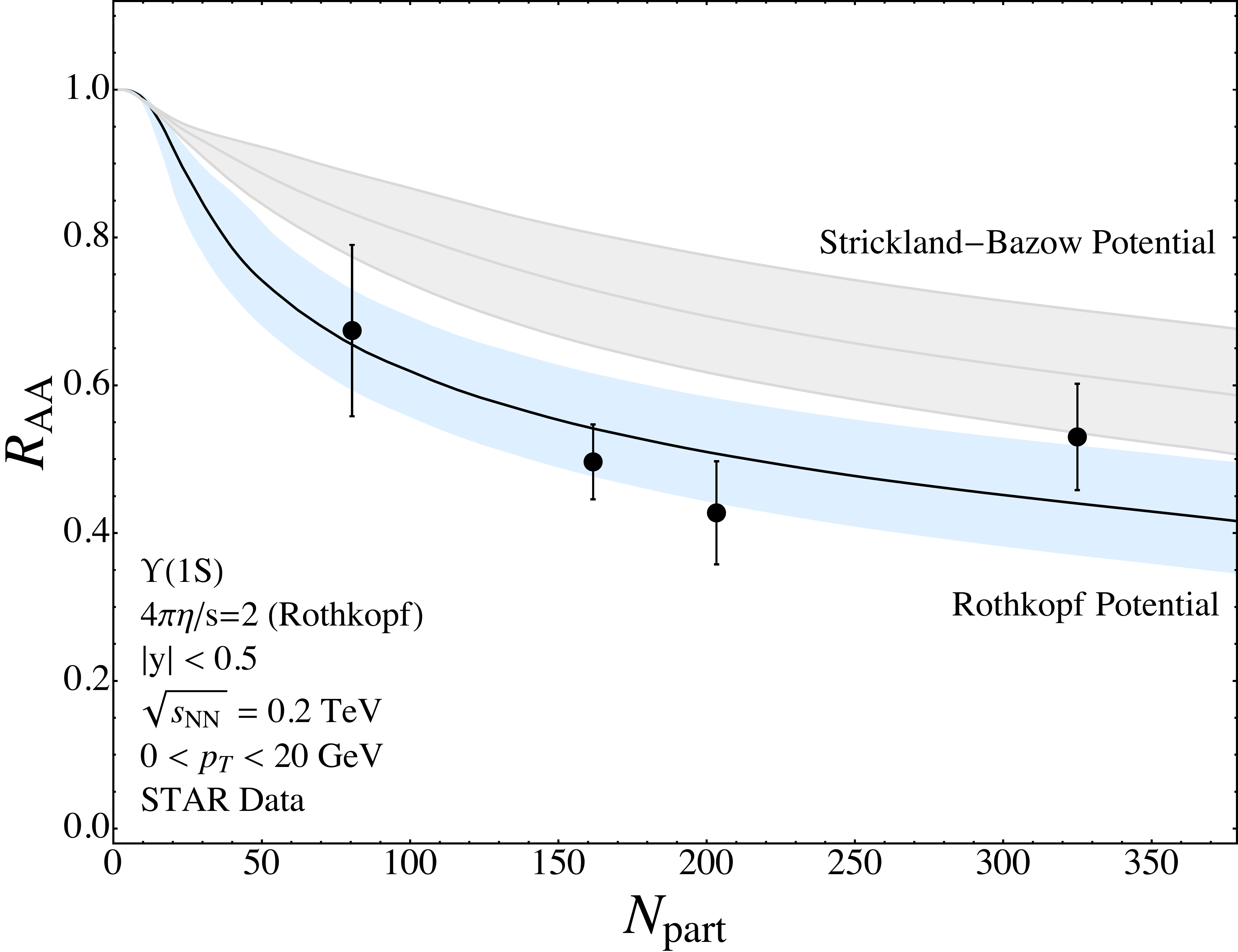}
}
\caption{(Left) The inclusive feed down suppression and regeneration results for the $\Upsilon(1S)$ bottomonium state as a function of $N_{\text{part}}$. (Right) A comparison between the Strickland-Bazow and Rothkopf potentials the inclusive feed down $\Upsilon(1S)$ at RHIC energies. Experimental $\Upsilon(1S)$ suppression data is taken from the STAR collaboration \cite{Ye:2017vuw}. 
}
\label{fig:regen}
\end{figure}

\begin{figure}[t!]
\centerline{
\includegraphics[width=0.503\linewidth, trim=0 0.27in 0 0]{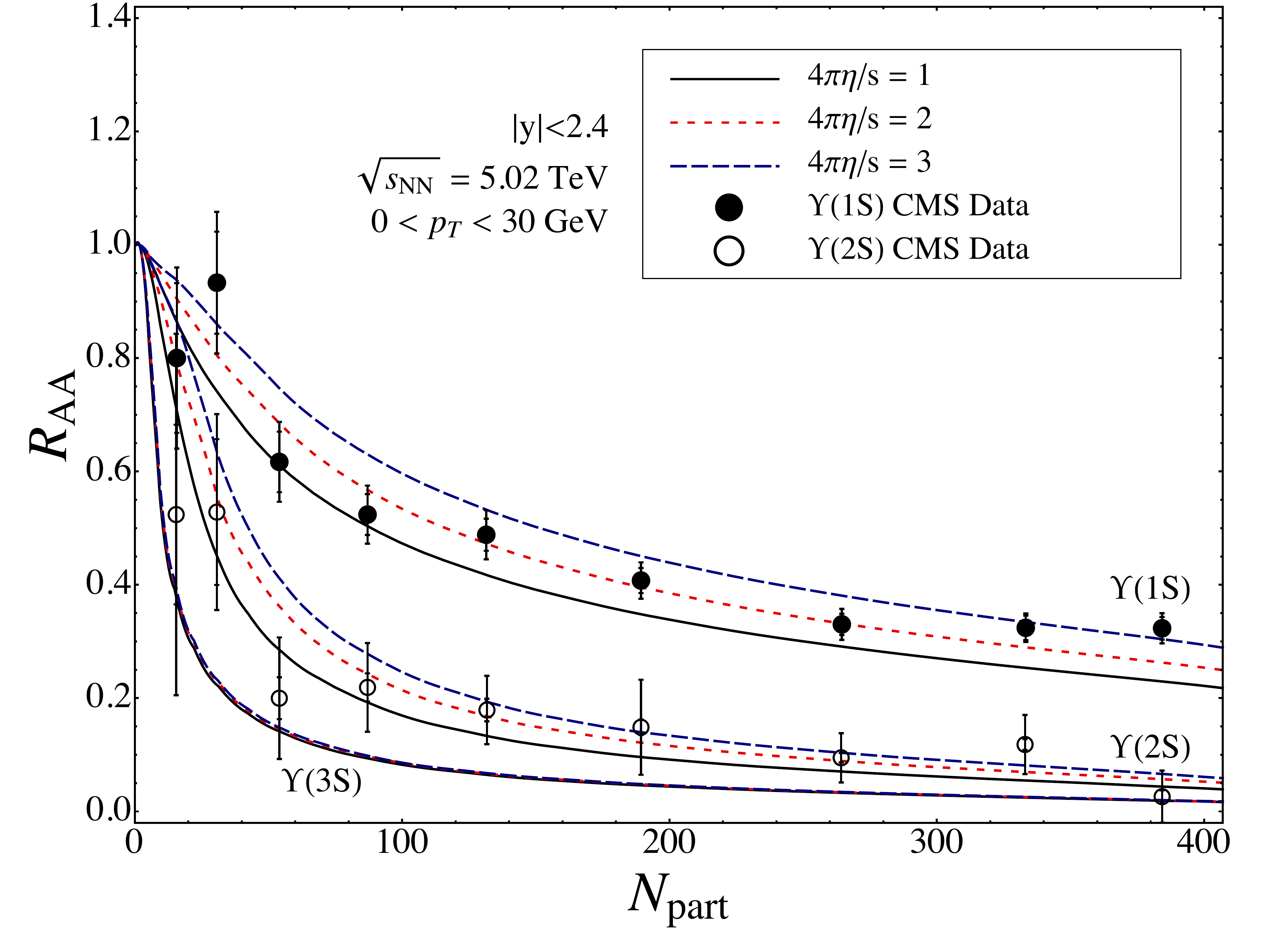}\hspace{2mm}
\includegraphics[width=0.485\linewidth]{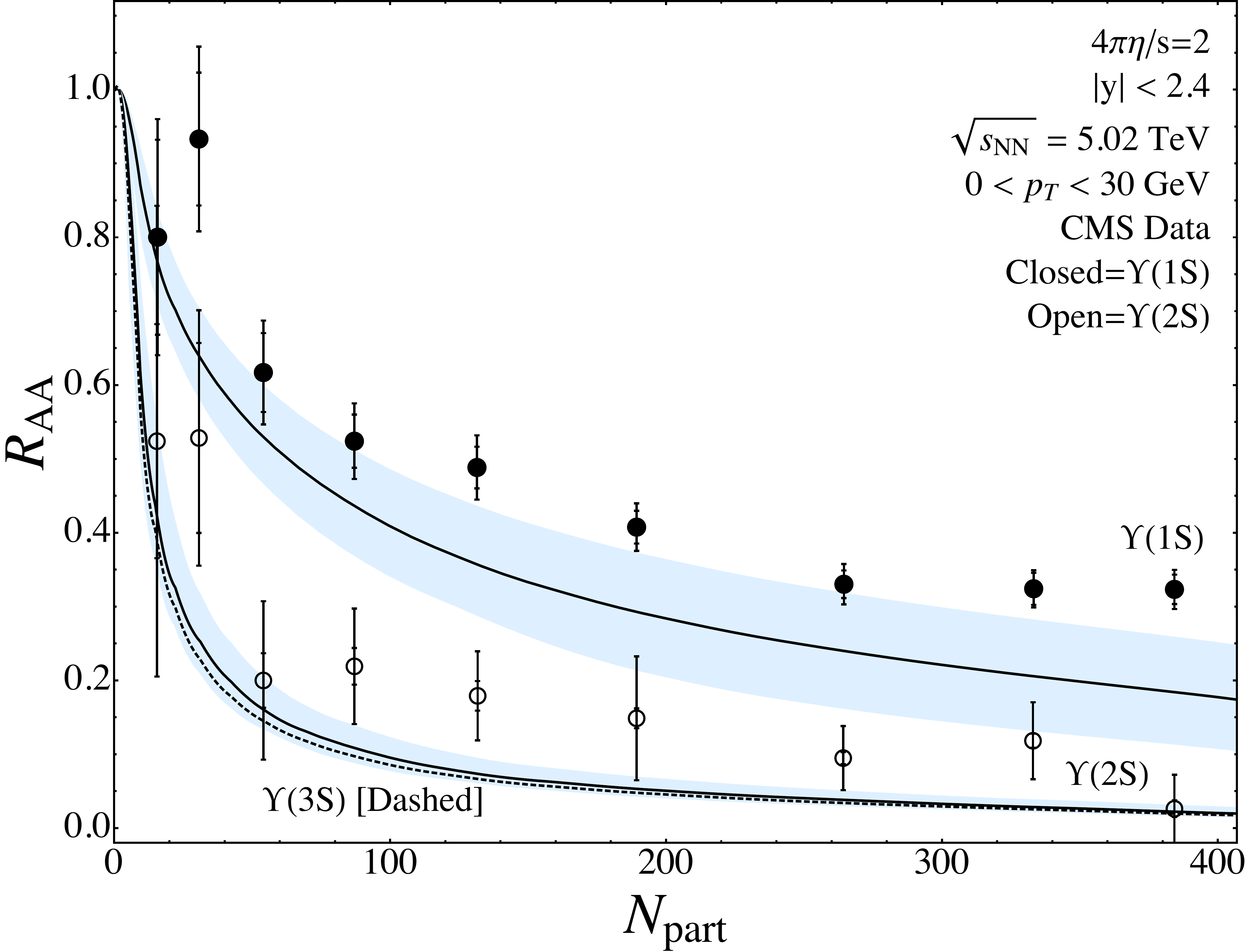}
}
\caption{(Left) The Strickland-Bazow bottomonium and (Right) Rothkopf bottomonium calculation for the inclusive $\Upsilon(nS)$ curves as a function of $N_{\text{part}}$. A comparison is made for the CMS experiment at the LHC. CMS data taken from \cite{Sirunyan:2018nsz}.
}
\label{fig:npart}
\end{figure}

\begin{figure}[t!]
\centerline{
\includegraphics[width=0.502\linewidth, trim=0 0.34in 0 0]{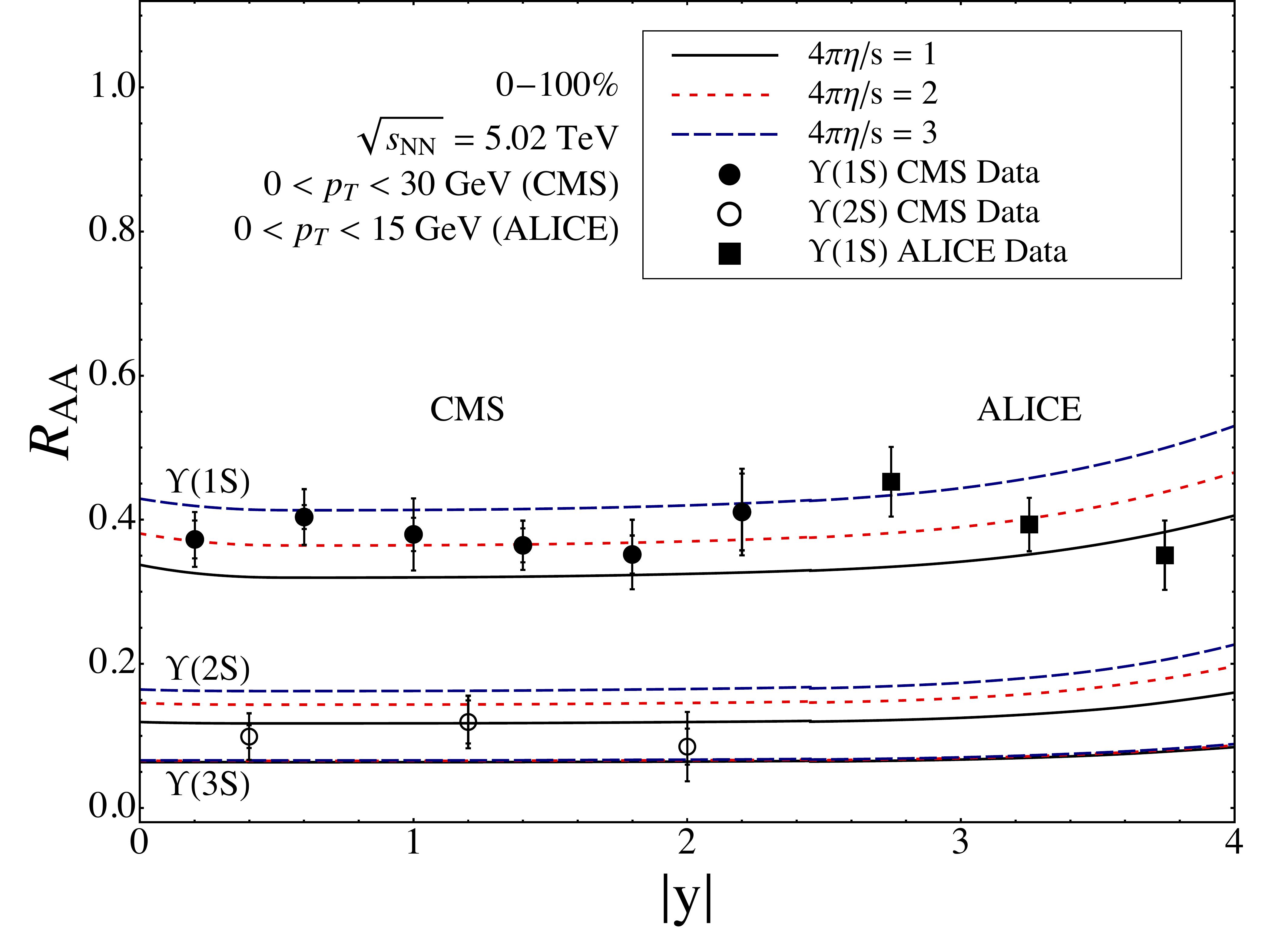}\hspace{2mm}
\includegraphics[width=0.485\linewidth]{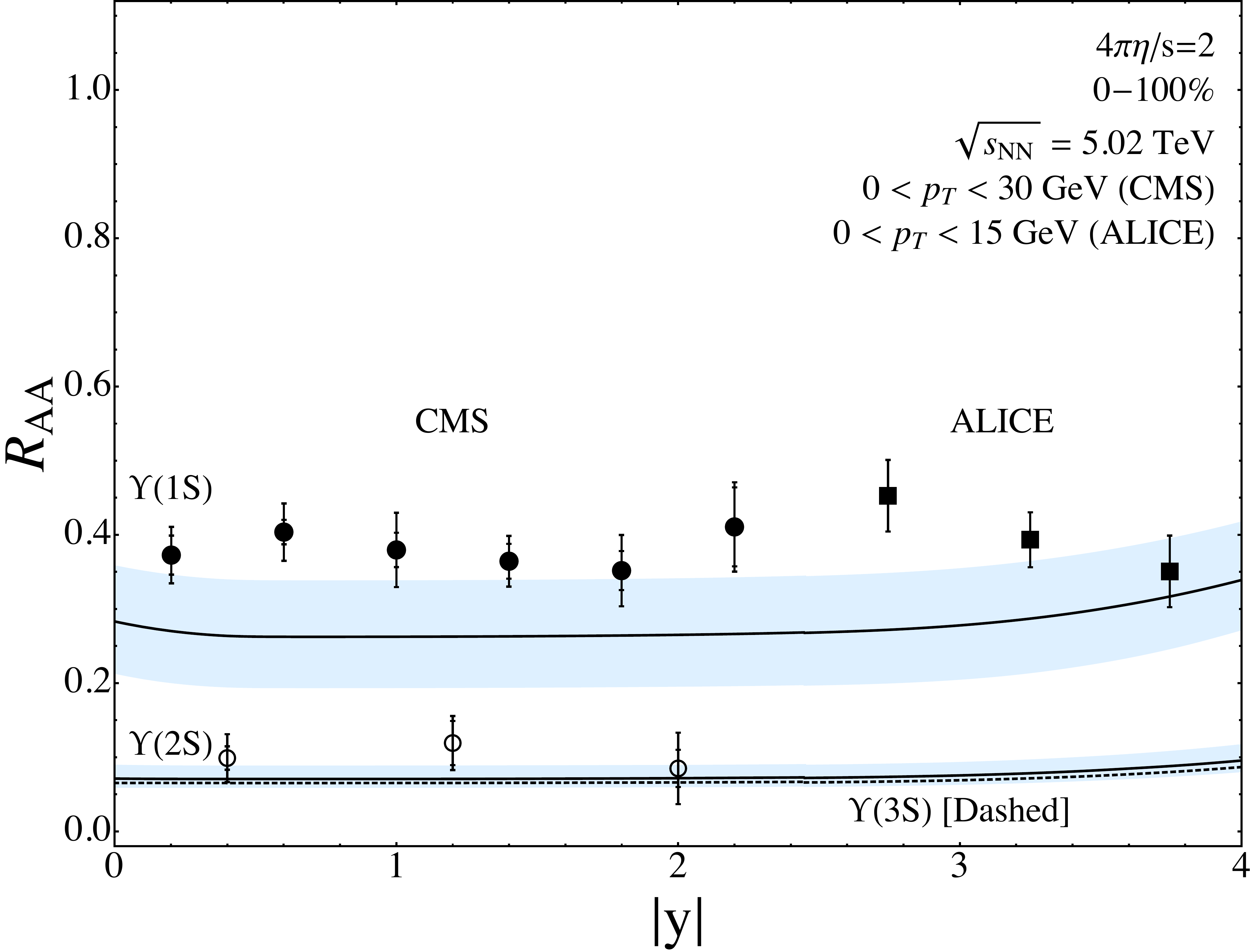}
}
\caption{(Left) The Strickland-Bazow potential and (Right) Rothkopf bottomonium calculation for the inclusive $\Upsilon(nS)$ $R_{AA}$ as a function of rapidity. A comparison is made for the CMS and ALICE experiments at the LHC. CMS data taken from \cite{Sirunyan:2018nsz} and ALICE data taken from \cite{Acharya:2018mni}.
}
\label{fig:rap}
\end{figure}

\section{Discussion and conclusions}
\label{sec:conclusions}
The model presented herein provides a more comprehensive description of bottomonium suppression seen at RHIC and LHC. The pNRQCD-based Strickland-Bazow potential describes URHICs at the LHC well while underestimating the amount of suppression seen at RHIC 200 GeV/nucleon collisions. The lattice-vetted Rothkopf potential does a good job at reproducing suppression in 200 GeV/nucleon RHIC Au-Au collisions, however slightly overestimates suppression for 2.76 TeV/nucleon and 5.02 TeV/nucleon LHC Pb-Pb collisions. Future work on continuum-extrapolated potential extractions using lattice QCD may relieve some tension with the data. In addition, the momentum-space anisotropy of the QGP causes one to insert non-equilibrium effects by hand to the lattice-QCD-based Rothkopf potential. Further advancements in the aHydro framework have pushed our understanding of URHICs for a variety of other observables such as particle spectra, flow coefficients, and HBT radii at both RHIC for 200 GeV/nucleon collisions \cite{Almaalol:2018gjh} and the LHC for 2.76 TeV/nucleon collisions \cite{Alqahtani:2017jwl}. It will be interesting to see if these news have an effect on the predicted bottomonium suppression

\vspace{3mm}

\noindent
{\bf Acknowledgments}:  B.~Krouppa and M.~Strickland were supported by U.S. DOE Award No.~DE-SC0013470.  A.~Rothkopf was supported by the DFG funded collaborative research center SFB 1225 ``IsoQUANT.''

\vspace{-2mm}






\bibliographystyle{elsarticle-num}
\bibliography{qm2018}







\end{document}